----------------------------- Start of body part 2

\documentstyle[12pt]{article}

\begin{document}

\begin{titlepage}
\title{Free Fermionic Elliptic Reflection Matrices and Quantum
Group Invariance}

\author{R. Cuerno \\
{\it Instituto de Matem\'aticas y F\'{\i}sica Fundamental, CSIC} \\
{\it Serrano 123, E--28006 Madrid, SPAIN}
\and
A. Gonz\'alez--Ruiz\thanks{Permanent address: Departamento
de F\'{\i}sica Te\'orica, Universidad Complutense, 28040 Madrid,
SPAIN} \\
{\it L.P.T.H.E.} \\
{\it Tour 16, 1er \'etage, Universit\'e Paris VI} \\
{\it 4 Place Jussieu, 75252 Paris cedex 05, FRANCE}}

\date{}

\maketitle

\begin{abstract}
Diagonal solutions for the reflection matrices
associated to the elliptic $R$ matrix of the eight vertex free
fermion model are presented. They lead through the second
derivative of the open chain transfer matrix to an XY
hamiltonian in a magnetic field which is invariant under a
quantum deformed Clifford--Hopf algebra.
\end{abstract}

\vskip-16.0cm
\rightline{{\bf IMAFF 93/11}}
\rightline{{\bf LPTHE--PAR 93/21}}
\rightline{{\bf April 1993}}
\vskip2cm

\end{titlepage}

\indent
A subject which is focusing great attention in Mathematical
Physics is to find a quantum group--like structure in which Baxter's
zero field 8 vertex matrix \cite{B} could act as an intertwiner
(for a recent attempt see \cite{U}). Recently a solution has
been found for the related case of the free--fermionic elliptic eight
vertex matrix introduced in \cite{F,BS}: it appears as an
intertwiner for the affinization of a quantum Hopf deformation
of the Clifford algebra in two dimensions,
$\widehat{CH_q(2)}$ \cite{CGLS}. In the
trigonometric limit this free--fermionic $R$ matrix becomes the
intertwiner of the quantum algebra ${\cal
U}_{\hat{q}}(\widehat{gl(1,1)})$ \cite{KS,RS}, or can also be
seen as the intertwiner for non classical nilpotent irreducible
representations of ${\cal U}_q(\widehat{sl(2)})$ for $q^4=1$ and
$\hat{q}=\lambda$, with $\lambda^2$ the eigenvalue of the
casimir $K^2$ \cite{M,Mar}. An integrable open chain hamiltonian can be
constructed for this trigonometric matrix \cite{Ch,S} which is
invariant under the ${\cal U}_q(sl(2))$ algebra in nilpotent
irreps \cite{AR} (also, see \cite{CGS} for the analogous $q^3=1$ case).

In this letter we consider the open chain hamiltonian associated with
the elliptic free fermionic matrix $R(u)$ (\ref{4.1}) \cite{F,BS},
$u$ being the complex spectral parameter living on a torus. Elliptic $K$
matrices will appear as solutions to the associated reflection
equations \cite{Ch,S}, and analogously to the trigonometric case
the associated open chain hamiltonian will be invariant under
the non--affine subalgebra of the bigger structure for which the
$R(u)$ matrix acts as an intertwiner.
First we will introduce the relevant material on
$\widehat{CH_q(2)}$ and $CH_q(2)$, and after working out
the trigonometric (nilpotent) case to fix notations
and procedure we will
consider the elliptic one: by the use of the $K$ matrices an
integrable open
chain hamiltonian is constructed which is
invariant under the quantum algebra $CH_q(2)$. It is no other
but an XY model in a magnetic field.
See \cite{HR} for a different approach to its quantum group
invariance.
In all these free fermionic models (both in the
trigonometric and in the elliptic cases) an analogous result is
obtained: ${\rm Tr} K_+(0)=0$. It forces us to define the
hamiltonian of the associated spin model through the second
derivative of Sklyanin's \cite{S} open chain transfer matrix;
only nearest-neighbour spins turn out to be coupled in contrast
to what could be expected \cite{L}. Some final comments are also included.

The most general solution for the elliptic free fermionic 8V
$R$ matrix is, in the parametrization of ref. \cite{BS}:
\begin{eqnarray}
R_{00}^{00}&=&1-e(u)e(\psi_1)e(\psi_2) \nonumber \\
R_{11}^{11}&=&e(u)-e(\psi_1)e(\psi_2) \nonumber \\
R_{01}^{01}&=&e(\psi_2)-e(u)e(\psi_1) \nonumber \\
R_{10}^{10}&=&e(\psi_1)-e(u)e(\psi_2) \label{4.1} \\
R_{01}^{10}&=&R_{10}^{01}=(e(\psi_1){\rm sn}(\psi_1))^{1/2}
(e(\psi_2){\rm sn}(\psi_2))^{1/2}(1-e(u))/{\rm sn}(u/2) \nonumber \\
R_{00}^{11}&=&R_{11}^{00}=-{\rm i}k(e(\psi_1){\rm sn}(\psi_1))^{1/2}
(e(\psi_2){\rm sn}(\psi_2))^{1/2}(1+e(u)){\rm sn}(u/2) \nonumber
\end{eqnarray}

\noindent
with $e(u)$ the elliptic exponential:
\begin{equation}
e(u)={\rm cn}(u)+{\rm i}\;{\rm sn}(u) \nonumber
\end{equation}

\noindent
$k$ the elliptic modulus and ${\rm cn}(u)$, ${\rm sn}(u)$ the Jacobi
elliptic functions of modulus $k$. In the sequel we will set
\begin{equation}
\psi_1=\psi_2 \equiv \psi \nonumber
\end{equation}

The quantum deformed Clifford--Hopf algebra of
dimension $D=2$, $CH_q(2)$ \cite{CGLS}, is generated by $\Gamma_{\mu}
\;(\mu=x,y),\; \Gamma_{3}$ and the central elements
$E_{\mu} \; (\mu=x,y)$ satisfying the following relations:
\begin{eqnarray}
& & \Gamma_{\mu}^2 = \frac{q^{E_{\mu}}-q^{-E_{\mu}}}{q-q^{-1}} \;\; , \;\;
\Gamma_{3}^2 = {\bf 1} \nonumber \\
& & \{ \Gamma_{x}, \Gamma_{y} \} =0 \;\; , \;\; \{\Gamma_{\mu},
\Gamma_{3} \} =0 \label{r} \\
& & [ E_{\mu}, \Gamma_{\nu} ] = [ E_{\mu}, \Gamma_{3} ] =
[ E_{\mu}, E_{\nu} ] = 0 \;\; \forall \mu, \nu \nonumber
\end{eqnarray}

\noindent
It is a Hopf algebra with the following
comultiplication $\Delta$, antipode $S$ and counit $\epsilon$:
\begin{equation}
\begin{array}{lll}
\Delta (E_{\mu}) = E_{\mu} \otimes {\bf 1} + {\bf 1} \otimes
E_{\mu}, & S(E_{\mu}) = -E_{\mu}, &\epsilon(E_{\mu}) = 0 \\
\Delta (\Gamma_{\mu}) = \Gamma_{\mu} \otimes q^{-E_{\mu}/2} +
q^{E_{\mu}/2} \Gamma_{3}
 \otimes \Gamma_{\mu}, & S(\Gamma_{\mu}) = \Gamma_{\mu}
\Gamma_{3}, & \epsilon(\Gamma_{\mu}) = 0 \\
\Delta(\Gamma_{3}) = \Gamma_{3} \otimes \Gamma_{3}, &
S(\Gamma_{3}) = \Gamma_{3}, & \epsilon(\Gamma_{3}) = 1 \\
\end{array}
\label{d}
\end{equation}

\noindent
A two dimensional representation is labelled by two complex
parameters $\xi = (\lambda_x, \lambda_y)$ and is
given by
\begin{eqnarray}
\pi_{\xi}(\Gamma_x) & = & \left(
\frac{\lambda_x-\lambda_x^{-1}}{q-q^{-1}} \right)^{1/2} \sigma_x
\;\; , \;\; \pi_{\xi}(\Gamma_y) = \left(
\frac{\lambda_y-\lambda_y^{-1}}{q-q^{-1}} \right)^{1/2} \sigma_y
\nonumber \\
\pi_{\xi}(\Gamma_3) & = & \sigma_z \;\;,\;\; \pi_{\xi}(q^{E_x}) = \lambda_x
\;\; , \;\; \pi_{\xi}(q^{E_y}) = \lambda_y \nonumber
\end{eqnarray}

\noindent
where $\{\sigma_j \}_{j=x,y,z}$ are Pauli matrices.
The (sort of) affine extension of this algebra, $\widehat{CH_q(2)}$, is
generated by $E_{\mu}^{(i)}, \Gamma_{\mu}^{(i)},
\;\;(i=0,1)$ and $\Gamma_{3} $ satisfying (\ref{r}) and (\ref{d})
for each value of $i$. Now a two dimensional irrep $\pi_{\xi}$ of
$\widehat{CH_q(2)}$
is labelled by three complex parameters $\xi = (z,\lambda_x,\lambda_y) \in
C_{\times}^3$ which fulfill the following constraint:
\begin{equation}
\frac{2 (\lambda_x - \lambda_y)}{(1-\lambda_x^2)^{1/2}
(1-\lambda_y^2)^{1/2}(z^2 - z^{-2})} = k
\label{4.2}
\end{equation}

\noindent
See more details in \cite{CGLS}. There (\ref{4.1}) is shown
to be the intertwiner between two irreps which satisfy
(\ref{4.2}) for the same value of $k$, with:
\begin{eqnarray}
z^2 & = & {\rm cn}(\varphi) + {\rm i}\;{\rm  sn}(\varphi) \nonumber \\
\lambda_x & = & \tanh x \;\; , \;\; \lambda_y = \tanh y
\nonumber \\
\tanh \left( \frac{x+y}{2} \right) & = & {\rm cn}(\psi) +
{\rm i}\;{\rm sn}(\psi) \nonumber \\
u & = & \varphi_1 - \varphi_2 \nonumber
\end{eqnarray}

\noindent
i.e. one has
\begin{equation}
\check{R}_{\xi_1 \xi_2} \Delta_{\xi_1 \xi_2}(a) = \Delta_{\xi_2
\xi_1}(a) {\check R}_{\xi_1 \xi_2} \;\;\; \forall a \in \widehat{CH_q(2)}
\nonumber
\end{equation}

\noindent
for $\check{R}(u) \equiv {\cal P} R(u)$, ${\cal P}$ being the
permutation matrix.

Consider now the trigonometric limit of the elliptic matrix
(\ref{4.1}). It can be written as
\begin{equation}
R(\theta) = \frac{1}{\lambda - \lambda^{-1}} \left( \begin{array}{cccc}
e^{\theta} \lambda - e^{-\theta} \lambda^{-1} & & & \\
 & e^{\theta} - e^{-\theta} & \lambda - \lambda^{-1} & \\
 & \lambda - \lambda^{-1} & e^{\theta} - e^{-\theta} & \\
 & & & e^{-\theta} \lambda - e^{\theta} \lambda^{-1} \\
\end{array} \right)
\label{2.1}
\end{equation}

\noindent
with $\lambda \equiv e^{{\rm i} \psi}$, $\theta \equiv
\frac{{\rm i}}{2} u$.
All Skylanin's requirements on symmetries (see \cite{S}
and (\ref{5.1})), are fulfilled with anisotropy fixed to $\eta = \pi / 2$. The
reflection--factorization equations for
diagonal $K_{\pm}(\theta)$ matrices, which are the ones
considered in this letter, are:
\begin{eqnarray}
& & R_{12}(\theta_1- \theta_2) \stackrel{1}{K}_-(\theta_1)
R_{12}(\theta_1+ \theta_2) \stackrel{2}{K}_-(\theta_2) =
\nonumber \\
& & = \stackrel{2}{K}_-(\theta_2) R_{12}(\theta_1+ \theta_2)
\stackrel{1}{K}_-(\theta_1) R_{12}(\theta_1- \theta_2)
\label{2.3} \\
& & R_{12}(-\theta_1+ \theta_2) \stackrel{1}{K}_+(\theta_1)
R_{12}(-\theta_1- \theta_2 - 2 \eta) \stackrel{2}{K}_+(\theta_2)
= \nonumber \\
& & = \stackrel{2}{K}_+(\theta_2) R_{12}(-\theta_1- \theta_2- 2 \eta)
\stackrel{1}{K}_+(\theta_1) R_{12}(-\theta_1+ \theta_2)
\label{2.4}
\end{eqnarray}

\noindent
Cherednik's solutions to these in the case of the 6V model are
also valid for (\ref{2.1}) with the mentioned requirement that $\eta=
\pi/2$. This implies in particular that
\begin{equation}
{\rm Tr}K_+(0)=0 \label{2.2}
\end{equation}

\noindent
which as anticipated makes it troublesome to define the open
chain hamiltonian through the standard formula \cite{S}:
\begin{equation}
H = \sum_{j=1}^{N-1} H_{j,j+1} + \frac{1}{2}
\stackrel{1}{K}'_-(0) + \frac{{\rm Tr}_0 \stackrel{0}{K}_+(0)
H_{N0}}{{\rm Tr} K_+(0)} \nonumber \\
\end{equation}
\begin{equation}
H_{j,j+1} \equiv {\cal P}_{j,j+1} \left. \frac{d R(\theta)}{d
\theta}\right|_{\theta =0}
\nonumber
\end{equation}

\noindent
However, using (\ref{2.2}) and the fact that
\begin{equation}
{\rm Tr}_0 \stackrel{0}{K}_+(0) H_{N0} = A \;\;{\bf 1}
\label{A}
\end{equation}

\noindent
where $A$ is a constant, we can see that there
still exists a well
defined open chain hamiltonian defining it as \cite{CGS}:
\begin{eqnarray}
H & \equiv & \frac{t''(0)}{4(T +2A)} = \nonumber \\
& = & \sum_{j=1}^{N-1} H_{j,j+1} + \frac{1}{2}
\stackrel{1}{K}'_-(0) + \frac{1}{2(T+2A)} \times \label{3.2} \\
& & \left\{ {\rm Tr}_0(\stackrel{0}{K}_+(0) G_{N0}) + 2{\rm Tr}_0
(\stackrel{0}{K}'_+(0) H_{N0}) + {\rm Tr}_0(\stackrel{0}{K}_+(0)
H_{N0}^2) \right\} \nonumber
\end{eqnarray}

\noindent
where $t(u)$ is Sklyanin's open chain transfer matrix, and:
\begin{eqnarray}
T & \equiv & {\rm Tr} K_+'(0) \nonumber \\
G_{j,j+1} & \equiv & {\cal P}_{j,j+1} \left. \frac{d^2 R_{j,j+1}(\theta)}{d
\theta^2} \right|_{\theta=0} \nonumber
\end{eqnarray}

\noindent
For the case of (\ref{2.1}) one gets the following hamiltonians:
\begin{equation}
H = \frac{1}{2} \sum_{j=1}^{N-1} \left\{ \sigma_{j}^x
\sigma_{j+1}^x + \sigma_{j}^y \sigma_{j+1}^y + \frac{\lambda +
\lambda^{-1}}{2} (\sigma_{j}^z + \sigma_{j+1}^z) \right\} \pm
\frac{1}{4} (\lambda-\lambda^{-1}) (\sigma_{1}^z-\sigma_{N}^z) \label{3.1}
\end{equation}

\noindent
where we have taken the standard limits $\alpha_{\pm}
\rightarrow \pm \infty$ in the arbitrary parameters appearing in
the $K_{\pm}(\theta)$ matrices \cite{DV}. This leaves us with an XX
model in a magnetic field which is invariant under ${\cal
U}_q(sl(2))$ transformations in nilpotent irrep $\lambda$ for
$q^4=1$, i.e.
\begin{equation}
[ H, \Delta^{(N)}(g) ] = 0  \label{comm}
\end{equation}

\noindent
with $g$ any generator of ${\cal U}_q(sl(2))$ in the nilpotent
irrep $\lambda$ for $q^4=1$ (see \cite{BGS} for a treatment of
general nilpotent closed chain spin models) and $\Delta^{(N)}$ the
comultiplication applied $N$ times (for an $N$ site chain).

We turn now our attention to the elliptic $R$ matrix
(\ref{4.1}). Taking $\psi_1 = \psi_2 = \psi$ it can be seen to have
the following set of symmetries:
\begin{eqnarray}
R(0) & = & (1-e(\psi)^2) \; {\cal P} \nonumber \\
{\cal P} R(u) {\cal P} & = & R(u) = R^{t_{12}}(u) \label{5.1} \\
R(u) R(-u) & = & \rho(u) \; {\bf 1} \nonumber \\
R^{t_1}(u) R^{t_1}(-u + 4 K) & = & \tilde{\rho}(u) \; {\bf 1}
\nonumber
\end{eqnarray}

\noindent
with $\rho(u)$, $\tilde{\rho}(u)$ some unimportant scalar
functions and $K$ the complete elliptic integral of the first
kind of modulus $k$. The most general diagonal solutions to the
corresponding reflection equations (\ref{2.3}),
(\ref{2.4}) are:
\begin{eqnarray}
K_-(u) & = & \left( \begin{array}{cc} {\rm cn}(\frac{u}{2})
{\rm dn}(\frac{u}{2}) \pm {\rm i} k' {\rm sn}(\frac{u}{2}) & \\
 & {\rm cn}(\frac{u}{2}) {\rm dn}(\frac{u}{2}) \mp {\rm i} k'
{\rm sn}(\frac{u}{2}) \\ \end{array} \right) \nonumber \\
K_+(u) & = & \left( \begin{array}{cc} k'^2
\frac{{\rm sn}(\frac{u}{2})}{{\rm dn}^2(\frac{u}{2})} \pm {\rm i} k'
\frac{{\rm cn}(\frac{u}{2})}{{\rm dn}(\frac{u}{2})} & \\
 & k'^2
\frac{{\rm sn}(\frac{u}{2})}{{\rm dn}^2(\frac{u}{2})} \mp {\rm i} k'
\frac{{\rm cn}(\frac{u}{2})}{{\rm dn}(\frac{u}{2})} \\ \end{array}
\right) \nonumber
\end{eqnarray}

\noindent
where $k'$ is the complementary elliptic modulus $k^2+k'^2=1$,
and we have used that $K_+(u) = K_-(-u+2K)$ solves (\ref{2.4}).
Note that there is no dependence on arbitrary parameters. This
seems to be the general case for elliptic $R$ matrices; a
dependence on an arbitrary parameter is expected, however, for
non diagonal $K$ matrices \cite{DG}.
Moreover the trigonometric limit of these matrices give us the
ones that yield the quantum group invariant hamiltonian
(\ref{3.1}). Note that we again have (\ref{2.2}) and (\ref{A}) even at the
elliptic level. So, whereas the symmetry properties (\ref{5.1})
enable us to construct the transfer matrix of an integrable open
chain model \`a la Sklyanin \cite{S}, we again have to resort to
(\ref{3.2}) to construct the corresponding spin hamiltonian. In
this case the solutions are given by
\begin{equation}
H= \frac{1}{2} \sum_{j=1}^{N} \left\{ (1+ \Gamma) \sigma_j^x \sigma_{j+1}^x +
(1- \Gamma) \sigma_j^y \sigma_{j+1}^y + h(\sigma_j^z +
\sigma_{j+1}^z) \right\} \pm \frac{{\rm i} k'{\rm sn}(\psi)}{2}
(\sigma_1^z - \sigma_N^z)
\label{6.1}
\end{equation}

\noindent
where $h= {\rm cn}(\psi)$, and $\Gamma= k \;{\rm sn}(\psi)$. (We have dropped
in (\ref{6.1}) a term proportional to the identity operator).
(\ref{6.1}) gives (\ref{3.1}) in the trigonometric limit. The
hamiltonians in (\ref{6.1}) are invariant under $CH_q(2)$
transformations in representations labelled by
\begin{equation}
\lambda_x = \frac{{\rm cn}(\psi) \pm {\rm i} k'{\rm sn}(\psi)}{1
+ k {\rm sn}(\psi)} \;\; , \;\; \lambda_y = \frac{{\rm cn}(\psi)
\pm {\rm i} k'{\rm sn}(\psi)}{1- k {\rm sn}(\psi)}
\end{equation}

\noindent
i.e. they fulfill commutation relations like (\ref{comm}) for
$g$ any generator of $CH_q(2)$ in the above representations.

We have shown that the $K$ matrices associated with the elliptic
free fermion $R$ matrix lead to an open chain hamiltonian which
is explicitly invariant under a quantum algebra. Investigating a
similar phenomenon would be interesting in the case of the zero
field 8V model. Also it would be interesting to examine the
implications of non--diagonal solutions for the $K$ matrices in
both cases. We have found here the appearance of the conditions
${\rm Tr}K_+(0)=0$ and ${\rm Tr}(\stackrel{0}{K}_+(0) H_{N0}) =
A \; {\bf 1}$ also at the elliptic level. In the trigonometric case
they can be interpreted in terms of quantum group representation
theoretical terms \cite{MN1,C}. It would be interesting to find
the explanation for the elliptic occurrence of the same
conditions. In this connection it would also be of interest to
find some ``gauge transformation" which would gauge
out the $u$ dependence from the elliptic $K$ matrices
\cite{MN2}, while taking $R(u)$ (\ref{4.1}) to a baxterized form \cite{J}.
Finally the Algebraic Bethe Anstaz \cite{S} for the elliptic
open chain described here would also be of interest, as well as
a detailed study of the algebraic ``contraction" involved in the
trigonometric limit $CH_q(2) \rightarrow {\cal U}_q(gl(1,1))$.
We hope to report on these matters in future publications.

\vspace{1cm}
{\bf Acknowledgments}

RC is pleased to thank A. Berkovich, C.
G\'omez, E. L\'opez and G. Sierra for discussions and
encouragement. AGR thanks the CNRS
Univ. Paris VI for the kind hospitality extended to him and H.J.
de Vega for valuable discussions. The research of both authors
is supported by the Spanish Ministry of Education and Science
through predoctoral fellowships PN89--11798388 (RC) and AP90--02620085 (AGR).


\newpage
\begin{thebibliography}{99}

\bibitem{B} R.J.Baxter, Exactly solved models in statistical
mechanics, Academic Press, 1982.
\bibitem{U} D.B.Uglov, ``The Lie algebra of the $sl(2,{\cal
C})$--valued automorphic finctions on a torus", and ``The quantum bialgebra
associated with the eight--vertex R--matrix", SUNY preprints,
February 1993.
\bibitem{F} B.U.Felderhof, Physica {\bf 66} (1973) 279.
\bibitem{BS} V.V.Bazhanov and Yu.G.Stroganov, Teor. Mat. Fiz.
{\bf 62} (1985) 37.
\bibitem{CGLS} R.Cuerno, C.G\'omez, E.L\'opez and G.Sierra, ``The
Hidden Quantum Group of the 8--vertex Free Fermion Model:
q--Clifford Algebras", preprint IMAFF-2/93, to appear in Phys.
Lett. B.
\bibitem{KS} L.H.Kauffman and H.Saleur, Commun. Math. Phys. {\bf
141} (1991) 293.
\bibitem{RS} L.Rozansky and H.Saleur, Nucl. Phys. {\bf B376} (1992) 461.
\bibitem{M} J.Murakami, Int. Jour. Mod. Phys. {\bf A7}, Suppl. 1B
(1992) 765.
\bibitem{Mar} M.Ruiz--Altaba, Phys. Lett. {\bf B279} (1992) 326.
\bibitem{Ch} I.V.Cherednik, Teor. Mat. Fiz. {\bf 61} (1984) 35.
\bibitem{S} E.K.Sklyanin, Jour. Phys. A:Math. Gen. {\bf 21} (1988) 2375.
\bibitem{AR} D.Arnaudon and V.Rittenberg, ``Quantum Chains with
${\cal U}_q(sl(2))$ Symmetry and Unrestricted Representations",
preprint CERN-TH.6786/93, January 1993.
\bibitem{CGS} R.Cuerno, C.G\'omez and G.Sierra, Jour. Geom.
Phys. {\bf 11} (1993).
\bibitem{HR} H.Hinrichsen and V.Rittenberg, Phys. Lett. {\bf B275}
(1992) 350, and preprint BONN HE--93--02.
\bibitem{L} M.L\"uscher, Nucl. Phys. {\bf B117} (1976) 475.
\bibitem{DV} C.Destri and H.J.de Vega, Nucl. Phys. {\bf B374} (1992) 692.
\bibitem{BGS} A.Berkovich, C.G\'omez and G.Sierra, ``Spin-Anisotropy
Commensurable Chains: Quantum Group Symmetries and N=2 SUSY",
preprint IMAFF-93-1, January 1993.
\bibitem{DG} H.J.de Vega and A.Gonz\'alez--Ruiz, in preparation.
\bibitem{MN1} L.Mezincescu and R.I.Nepomechie, Int. Jour. Mod.
Phys. {\bf A6} (1991) 5231 and addendum Int. Jour. Mod. Phys.
{\bf A7} (1992) 5657.
\bibitem{C} R.Cuerno, in preparation.
\bibitem{MN2} L.Mezincescu and R.I.Nepomechie, Jour. Phys.
A:Math. Gen. {\bf 24} (1991) L17.
\bibitem{J} V.F.R.Jones, Int. Jour. Mod. Phys. {\bf A6} (1991) 2035.

\end{thebibliography}
\end{document}